\documentstyle[preprint,eqsecnum,aps]{revtex}

\begin{document}
\title{Fermi and Gamow-Teller Strength in Charge Exchange with Radioactive Beams }
\author{C. A. Bertulani$^a$ and P. Lotti$^b$}
\address{$^a$ Instituto de F\'\i sica, Universidade Federal \\
do Rio de Janeiro \\
21945-970 Rio de Janeiro, RJ, Brazil. E-mail: bertu@if.ufrj.br\\
$^b$ INFN, Sezione di Padova, via F. Marzolo 8, \\
I-35131 \ Padova, Italy}
\maketitle

\begin{abstract}
At forward angles, and bombarding energies $E>200$ MeV, the $(p,n)$ and $%
(n,p)$ reactions are thought to be directly proportional to the Gamow-Teller
transition strengths in the nuclei. Assuming that this relationship also
holds for charge exchange induced by high-energy heavy ions, it would be
very useful in studies with radioactive beams. Contrary to this expectation,
we show that the determination of Gamow-Teller and Fermi matrix elements
from heavy-ion charge-exchange at forward angles is very inaccurate.
\end{abstract}


A major experimental and theoretical effort is under way to study the
properties and applications of drip-line nuclei (for a review, see ref. \cite
{Nii94}). One of the main issues of this rapidly growing field is to
understand the basic features of many unstable nuclei, which can only be
assessed by using beams of radioactive nuclei. Of great relevance is the
study of spin-isospin properties in unstable nuclei. A well-defined
proportionality between heavy ion charge-exchange and beta-decay transition
strengths would have important consequences. Such a relationship would be
particularly useful to allow beta transition strength functions to be mapped
out in drip-line nuclei. An accurate link between heavy ion exchange
reactions and the well understood beta-decay processes would also put our
understanding of the nuclear structure and reaction mechanisms of unstable
nuclei on a surer quantitative footing.

It is well known that $(p,\ n)$ and $(n,\ p)$ charge-exchange reactions at
intermediate energies ($E>100$ MeV) have been a powerful spectroscopic tool
for measuring Gamow-Teller (GT) transition matrix elements \cite{Go80,Ta87}.
This approach can also be used in reactions with radioactive beams, using a
hydrogen gas target. However, due to the low luminosity of the beams, this
process yields low counting rates. A more promising procedure would be the
use of heavy ion charge exchange (HICEX) instead of $(p,\ n)$ collisions. In
ref. \cite{Os92} the possibility to use HICEX as a probe of Gamow-Teller
matrix elements was suggested and a relation between the forward cross
section and the Gamow-Teller matrix elements was found. A possible drawback
of this approach is the complication originated by the internal structure of
the target nucleus. This method was used as a guidance in ref. \cite{St96}
to draw conclusions on Gamow-Teller (and Fermi) matrix elements in $^{13}C$
and $^{13}N$ from the mirror nuclei reaction $^{13}C(^{13}N,\ ^{13}C)^{13}N$.

In this article we develop a strong absorption model of HICEX based on a
single particle approach. We use this simple structure model for qualitative
purposes only. Our aim is to access the reliability of the relationship of
HICEX measurements and the Gamow-Teller and Fermi matrix elements. The key
point to be discussed here is that if we cannot show the existence of a
valid relationship for such a simple model and for rather simple reaction
partners, then very probably it would not exist for more complicated nuclear
models. Because of the use of mirror nuclei, and their isospin symmetry, the
experimental data on the $^{13}C(^{13}N, \ ^{13}C)^{13}N$ reaction is also
an ideal test of the theory.

In heavy-ion charge exchange reactions it is important to know the
underlying nature of the exchange mechanism. In ref. \cite{Le89} it was
shown that in heavy ion collisions at high energies ($E>100$ MeV.A) the
exchange mechanism is dominated by the fundamental $\pi$- and $\rho$-
exchange. The differential cross section for the process is given by 
\begin{equation}
{\frac{d\sigma }{d\Omega}} = {\frac{k^{\prime}}{k}} \ \bigg( {\frac{\mu }{%
4\pi^2 \hbar ^2}}\bigg)^2 \ (2J_1+1)^{-1} (2J_2+1)^{-1} \ \sum_{\alpha_1, \
\alpha_2} \Bigg| \ T_{if}(\alpha_1, \ \alpha_2) \Bigg|^2 \ ,  \label{sig1}
\end{equation}
where $\mu$ is the reduced mass of the system (1+2), $k$ ($k^{\prime}$) is
the initial (final) center-of-mass momentum, $J_1$ and $J_2$ are the angular
momenta in the entrance channel, and the sum is an average over initial
angular momentum projections and a sum over final angular momentum
projections, denoted respectively by $\alpha_1$ and $\alpha_2$.

The transition matrix $T_{if}$ is given by \cite{Be93} 
\begin{equation}
T_{if}= {\frac{1}{(2\pi)^3}} \int d^3 R \ \chi^{(-)*}({\bf R}) \ \chi^{(+)}(%
{\bf R})\Big\} \int d^3q \ e^{i{\bf q.R}} \ {\cal M} (\alpha_1,\ \alpha_2, \ 
{\bf q})\ ,  \label{tmat}
\end{equation}
where ${\bf R}\equiv ({\bf b}, z)$ is the relative coordinate between the
nuclei, $\chi^{(-)}({\bf R})$ $\Big[\chi^{(+)}({\bf R})\Big]$ is the
incoming (outgoing) scattering wave obtained in terms of the optical
potential for elastic scattering, and 
\begin{equation}
{\cal M} = \int d^2 r_1 \ d^3 r_2 \ \delta \rho_1({\bf r_1}) \ \delta \rho_2
({\bf r_2}) \ e^{i{\bf q}.({\bf r_1-r_2})} \ V_{exch} ({\bf q})\ ,
\label{mmat}
\end{equation}
where $\delta\rho_i$ is the transition density of the nucleus $i$, which
includes spin, isospin and spatial transitions. The meson-exchange potential 
$V_{OBEP}$ is a sum of $\pi$-, $\rho$-, and $\omega$-exchange interaction.
We use a modified meson-exchange interaction to include a non-spin-flip
term: $V_{exch} = V_{OBEP} + V_{\tau} \tau_1.\tau_2$ \cite{Be93,Br81}.

The eq. \ref{mmat} is separable in terms of the spin and isospin variables,
yielding ${\cal M}={\cal M}_{12}+{\cal M}_{2 \leftrightarrow 1}$, where 
\begin{eqnarray}
{\cal M}_{12} &=& \int d^2 q \ \bigg\{ F_{\sigma\tau}({\bf q}) \
\sum_{\mu=1}^3 <\alpha_f(1)\Big|\tau_1^{(+)} \sigma_\mu \ e^{i{\bf q.r_1}}%
\Big| \alpha_i (1)> \ <\alpha_f(2)\Big|\tau_2^{(-)} \sigma_\mu \ e^{i{\bf %
q.r_2}}\Big| \alpha_i (2)>  \nonumber \\
&+& G_{\sigma\tau} ({\bf q}) \ \sum_{\mu,\mu^{\prime}=1}^3 \hat{q}_\mu \hat{q%
}_{\mu^{\prime}} <\alpha_f(1)\Big|\tau_1^{(+)} \sigma_\mu \ e^{i{\bf q.r_1}}%
\Big| \alpha_i(1)> \ <\alpha_f(2)\Big|\tau_2^{(-)} \sigma_{\mu^{\prime}} \
e^{i{\bf q.r_2}}\Big| \alpha_i(2)>  \nonumber \\
&+& H_\tau({\bf q}) \ <\alpha_f(1)\Big|\tau_1^{(+)}\ e^{i{\bf q.r_1}} \Big| %
\alpha_i(1)> \ <\alpha_f(2)\Big|\tau_2^{(-)}\ e^{i{\bf q.r_2}} \Big| %
\alpha_i(2)> \ .  \label{mmat2}
\end{eqnarray}
In the equation above the sums over $\mu$ and $\mu^{\prime}$ run over the
spherical components of the spin vector. $F_{\sigma\tau}({\bf q})$, $%
G_{\sigma\tau}({\bf q})$ and $H_\tau({\bf q})$ are functions of the central
(L=0) and the tensorial (L=2) component of the charge-exchange potential.

Since the initial and final wavefunctions are decomposed into their spatial
and spin-dependent parts, it is appropriate to expand the exponential
functions in eq. \ref{mmat2} into multipoles and use the Wigner-Eckart
theorem, to obtain sums over products of Clebsh-Gordan coefficients and
reduced matrix elements of the form $<\phi^{(f)}_{j^{\prime}}||j_I(qr)[%
\sigma\otimes Y_I]_{I^{\prime}}\ \tau^{(\pm)} ||\phi^{(i)}_j>$, $%
<\phi^{(f)}_{j^{\prime}}||j_I(qr)[\sigma\otimes Y_I]_{I^{\prime}}
||\phi^{(i)}_j>$, and $<\phi^{(f)}_{j^{\prime}}||j_I(qr)\ \tau^{(\pm)}
||\phi^{(i)}_j>$. These matrix elements can be calculated using eqs.
(A.2.19) and (A.2.24) of ref. \cite{La80}. They involve geometric
coefficients and overlap integrals of the form 
\begin{equation}
{\cal I}_{Ijj^{\prime}}(q)=\int R_{j^{\prime}} (r) \ R_j (r) \ j_I(qr) \
r^2\ dr \ ,  \label{i}
\end{equation}
in terms of the single-particle wavefunctions $R_j(r)$. The cross section
results from interference of terms corresponding to several possibilities of
angular momentum transfer.

From \ref{mmat2} it is straightforward to show the relationship between the
heavy-ion charge-exchange cross sections and the Gamow-Teller and Fermi
matrix elements. In beta-decay experiments a similar matrix element appears,
but with the exponential factors replaced by $\exp{i({\bf p}_e+{\bf p}_\nu).%
{\bf r}}$, where ${\bf p}_e$ and ${\bf p}_\nu$ are the electron and the
neutrino momentum, respectively. This exponential factor can be approximated
by the unity, if $|{\bf r}|<\hbar/p_e$ and $\hbar/p_\nu$. Since the energies
of the light particles involved in beta-decay are of the order of a few MeV,
the Compton wavelengths of the electron and the neutrino are approximately
equal to $10^3$ fm, i.e., about hundred times larger than the nuclear sizes
over which the integration occurs. This contrasts with charge exchange
experiments, where the exponential factor in \ref{mmat2} involves the
Compton wavelength of $\pi$ and $\rho$. Thus, the relation ${\bf q.r}\ll 1$
is not well justified for HICEX.

Assuming that the ``low-$q$" (or long-wavelength) approximation is also
valid for HICEX, i.e., inserting eq. \ref{mmat2}, with the exponentials
replaced by the unity in eqs. \ref{sig1}-\ref{mmat2}, one gets 
\begin{equation}
{\frac{d\sigma }{d\Omega}} \simeq c_1\ B_1(GT) \ B_2(GT) + c_2\ B_1(F) \
B_2(F) + c_3 \ \bigg[ B_1(F) \ B_2(F)\ B_1(GT) \ B_2(GT)\bigg]^{1/2} \ ,
\label{sig2}
\end{equation}
where 
\begin{equation}
B(GT)={\frac{1}{2j+1}}\ \sum_{\alpha^{\prime},\alpha} \ \Big|%
<\alpha^{\prime}|\sigma \tau^{(\pm)} |\alpha>\Big|^2 \ \ {\rm and} \ \ B(F)={%
\frac{1}{2j+1}}\ \sum_{\alpha^{\prime},\alpha} \ \Big|<\alpha^{\prime}|%
\tau^{(\pm)} |\alpha>\Big|^2  \label{bgt}
\end{equation}
are the Gamow-Teller and Fermi strengths, respectively. The coefficients $%
c_1 $, $c_2$, and $c_3$ are products of geometric factors and integrals over
the $q$- and $R$-coordinates. The ``low-$q$" approximation \ref{sig2} also
includes a sum over all single-particle transitions, which is not explicitly
shown for simplicity.

Eq. \ref{sig2} is the result that one would like to have. Assuming that the
charge-exchange potential (corrected for medium effects, etc.) is accurately
known and that the same is true for the scattering waves, then the
coefficients $c_i$ could be reliably calculated. An experimental measurement
of HICEX would then provide the magnitude of $B(GT)$'s and $B(F)$'s for
several nuclei. This would be a specially important technique for studies of
dripline nuclei. We now investigate if this relation is valid using the $%
^{13}C(^{13}N,\ ^{13}C)^{13}N$ reaction at 105 MeV/nucleon \cite{St96} for
comparison.

Following ref. \cite{Br81} we normalize the interaction so that at zero
momentum transfer $V_{OBEP}\equiv V_{\sigma\tau}=265$ MeV.fm$^3$ and $%
V_\tau=100$ MeV.fm$^3$, appropriate for proton laboratory energies of order
of 100 MeV. We describe the scattering waves in terms of eikonal
wavefunctions. Since an optical potential for this reaction at 100
MeV/nucleon is not known, we use an optical potential which fits the
reaction $^{12}C+^{12}C$ at 85 MeV/nucleon \cite{Bue82,Be93}. To calculate
the single-particle wavefunctions we use a potential of the form $%
V=V_0f(r)+V_{SO}({\bf l.\sigma})\ (df/dr)/r$, with $f(r)=\Big[1+\exp{%
(r-R_0)/a}\Big]^{-1}$, for the neutrons. For the protons we add the Coulomb
potential of a uniformly charged sphere of radius $R_0$. For both $^{13}C$
and $^{13}N$ we choose the set of parameters $V_0=-53$ MeV, $R_0=2.86$ fm, $%
a=0.65$ fm, and $V_{SO}=-15.5$ MeV. In this simple model, the relevant
transitions are $1p_{1/2}^\pi \longrightarrow 1p_{1/2}^{\pi^{\prime}}$ ($%
\pi=n,$ or $p$, i.e., one proton in $^{13}N$ goes to a neutron in $^{13}C$,
and vice-versa) involving the operator $\tau$ (Fermi transitions). For
Gamow-Teller transitions ($\sigma\tau$ operator) the relevant transitions
are $1p_{1/2}^\pi \longrightarrow 1p_{1/2}^{\pi^{\prime}}$ and $1p_{1/2}^\pi
\longrightarrow 1p_{3/2}^{\pi^{\prime}}$.

In table 1 we show the matrix elements of the single particle transitions: $%
B_{1/2^-} \equiv B_{1/2^-} \longrightarrow B_{1/2^-}$ and $B_{3/2^-} \equiv
B_{1/2^-} \longrightarrow B_{3/2^-}$ for Fermi and Gamow-Teller transitions.
For comparison, we also give the values of these matrix elements, in the
case that the overlap integral includes a spherical Bessel function of
zeroth-order, $j_0(qr)$, for additional two values of $q$. The matrix
elements are drastically reduced for large values of $q$, especially for $%
1p_{1/2} \longrightarrow 1p_{3/2}$ transitions.

In figure 1 we compare the experimental \cite{St96} and the theoretical
calculation of the differential charge exchange cross section for the double
excitation case, i.e. when both target ($^{13}C$) and projectile ($^{13}N$)
go to the $3/2^-$ state. Both experimental data and the calculation are
given in arbitrary units. Thus, only the angular dependence of the cross
section can be discussed. The angular dependence is rather insensitive to
the structure model for the nuclei, but strongly sensitive on the parameters
of the optical potential used to calculate the scattering wavefunctions.
Only when such parameters are known from, e.g., elastic scattering one can
have a better knowledge of the influence of the Gamow-Teller and Fermi
matrix elements on the charge exchange cross section. This, of course,
assuming that the exchange potential is well known. In fact, the form of the
exchange potential is crucial for this aim.

In figure 2 we plot the ratio between the exact calculation and the low-q
approximation, eq. \ref{sig2}. With the single particle basis the expansion
can be stopped at the $I=5$ term. In fact, the term with $I=0$ dominates.
This is shown in the figure where the contribution of the $I=0$ (dashed
curve) term of the expansion is shown. The total sum includes terms up to $%
I=5$ (solid curve). However, note that the ratio between the exact
calculation and the low-q approximation amounts to a factor of 5 at the
forward direction. This difference can be understood in a simple way by
using eq. \ref{tmat}. If strong absorption and final state interactions
could be neglected, $\chi ^{(-)*}({\bf R})\ \chi ^{(+)}({\bf R})\approx \exp %
\Big\{i{\bf Q.R}\Big\}$, where ${\bf Q=k^{\prime }-k}$ is the momentum
transfer to the c.m. scattering. Thus, the integral over ${\bf R}$ yields a
delta function, $\delta ({\bf q-Q})$. That is, $Q=0$ (forward scattering)
would imply $q=0$, too. Very forward scattering would then justify replacing
the exponentials in eq. \ref{mmat2} by unity, and an equation of the form of
eq. \ref{sig2} would be obtained. However, strong absorption modifies this
scenario: ${\bf Q}$ is not directly related to ${\bf q}$, and the equation 
\ref{sig2} is a bad approximation. The larger the absorption radius, the
worse this approximation should be. This may somewhat explain why for $(p,n)$
reactions the low-q approximation is better justified, since the absorption
radius is smaller in that case. But, even for $(p,n)$ reactions the low-q
approximation seems to be ill-defined, as was shown recently \cite{Au94},
especially for the weak transitions.

Another possibility to get direct information on the Fermi and Gamow-Teller
strengths from HICEX reactions would be in the case that the expansion of
the exponential function in \ref{mmat2} can be stopped at the term with $I=0$%
. One gets matrix elements of the form 
\begin{equation}
<\phi_{j^{\prime}l^{\prime}m^{\prime}}^{\pi^{\prime}} | \sigma_\mu \tau^\nu
j_0(qr)|\phi_{jlm}^\pi> =\hat{j} (j1m\mu|j^{\prime}m^{\prime}) {\cal F}%
_{jj^{\prime}} (q) \bigg[ B(GT)\bigg]^{1/2} \ ,  \label{approx2}
\end{equation}
and a similar expression for Fermi transitions. The function ${\cal F }%
(q)_{jj^{\prime}} = |{\cal I}_{0jj^{\prime}}(q)/{\cal I}_{0jj^{%
\prime}}(q=0)|^2$ has to be calculated with accuracy in this case. Similar
arguments are given in ref. \cite{Ta87} for $(p,n)$ reactions.

In figure 3 we show the percentage deviation between the exact calculation
and a calculation done with only the $I=0$ term of the expansion. Also shown
(dashed curve) is the differential cross section in arbitrary units. At $%
\theta =0$ the difference is about 2\% and increases largely close to the
minima of the cross section. This is because, close to the minima (which are
basically determined by the $I_0$ term) the spherical Bessel functions $%
j_{I>0}$ contribute more to the integrals.

Although the ($I=0$)-approximation is not bad at $\theta=0$ compared to the
full calculation, it cannot be considered as a useful tool to obtain the
Fermi and Gamow-Teller strengths from HICEX, unless a good model for the
overlap integrals can be done. This is however just what one would like to
extract from the experiments, and not from theory. As one sees from table 1
the matrix elements involved in HICEX are strongly modified by the short
range property of the exchange interaction. Only a small part of the
transition densities, close to the surface of the nucleus is relevant for
the matrix elements. In contrast, in beta-decay the whole transition density
is relevant. To show this we decrease artificially the masses of the pion
and the rho by a factor $\alpha$. These masses enter into the form of the
exchange interaction and are responsible for its range \cite{Br81,Be93}. As
we see from figure 4, as $\alpha$ increases, increasing the range of the
interaction, the ratio between the exact calculation and the low-q
approximation decreases, reaching unity at $\alpha\approx 10$. This means
that eq. \ref{sig2} would be valid for an interaction of range $r_0 \approx
10 \times (\hbar/m_\pi c) \approx 10$ fm.

In conclusion, we have shown that it is very difficult to obtain reliable
estimates of Gamow-Teller or of Fermi strength from HICEX experiments. This
might be surprising, since weak interaction strengths determined from $(p,n)$
reactions are astonishingly good, although they have been shown to be
inaccurate for important Gamow-Teller transitions whose strengths are a
small fraction of the sum rule limit \cite{Au94}.

\bigskip\bigskip
\noindent{\bf Acknowledgments}

\medskip

We acknowledge the support of the Brazilian funding agencies CAPES and
FUJB/UFRJ and by the GSI/Darmstadt. This work has been partially supported
by MCT/FINEP/CNPQ(PRONEX) under contract \# 41.96.0886.00.

\newpage
\bigskip
\begin{table}[tbp]
\caption{ Modified Fermi and Gamow-Teller strengths for single particle
transitions and as a function of $q$. The Gamow-Teller and Fermi operators
are replaced by their respective product with a spherical Bessel function $%
j_0(qr)$. For $q=0$, $j_0(qr)=1$, and the usual Fermi and Gamow-Teller
matrix elements are restored. }
\label{table:1}
\begin{center}
\begin{tabular}{|c|c|c|c|}
& $q=0$ & $q=0.5$ fm$^{-1}$ & $q=1$ fm$^{-1}$ \\ \hline
$B_{1/2^-}(F)$ & 1 & 0.65 & 0.14 \\ \hline
$B_{1/2^-}(GT)$ & 2.66 & 1.73 & 0.37 \\ \hline
$B_{3/2^-}(GT)$ & 0.33 & 0.18 & 0.02
\end{tabular}
\end{center}
\end{table}

\bigskip 
{\bf Figure Captions}\\

{\bf Fig. 1} - Differential cross section of the charge exchange reaction $%
^{13}C(^{13}N,^{13}C)^{13}C$ at 105 MeV/nucleon for the double excitation
case (target and projectile go to the $3/2^{-}$ state). Experimental data 
\cite{St96} and calculation are given in arbitrary units. \\{\bf Fig. 2} -
Ratio between the exact calculation and the approximation \ref{sig2} for the
charge exchange reaction $^{13}C(^{13}N,^{13}C)^{13}C$ at 105 MeV/nucleon,
as a function of the scattering angle. \\{\bf Fig. 3} - Percentage deviation
from the exact calculation and the approximation obtained by keeping only
the $I=0$ term of the multipole expansion \ref{mmat2} for the charge
exchange reaction $^{13}C(^{13}N,^{13}C)^{13}C$ at 105 MeV/nucleon, as a
function of the scattering angle. \\{\bf Fig. 4} - Ratio between the exact
calculation and the approximation \ref{sig2} as a function of an scaling
factor $\alpha $ for the charge exchange reaction $%
^{13}C(^{13}N,^{13}C)^{13}C$ at 105 MeV/nucleon. The scaling factor $\alpha $
is used to reduce the pion and the rho masses in the charge exchange
interaction.

\end{document}